\documentclass[11pt]{llncs}
\usepackage{amssymb,amsmath,enumerate,graphicx}
\usepackage{verbatim}

\usepackage{tikz}
\usetikzlibrary{snakes,shapes}
\tikzstyle{hyb}=[rectangle,draw,minimum size=3.5mm]
\tikzstyle{tre}=[circle,draw,minimum size=3.5mm]
\newcommand{\etq}[1]{%
\draw (#1) node {\footnotesize $#1$};
}

\newcommand{\pathgr}{\!\rightsquigarrow\!{}}

\renewcommand{\geq}{\geqslant}

\begin{document}
 
\title{All that Glisters is not Galled}

\author{Francesc Rossell\'o\inst{1} \and
Gabriel Valiente\inst{2}} 
\institute{Department of Mathematics and Computer Science, University
of the Balearic Islands, E-07122 Palma de Mallorca,
\texttt{cesc.rossello@uib.es} \and
Algorithms, Bioinformatics, Complexity and Formal Methods Research
Group, Technical University of Catalonia, E-08034 Barcelona,
\texttt{valiente@lsi.upc.edu} }
\maketitle

\begin{abstract}
Galled trees, evolutionary networks with isolated reticulation cycles,
have appeared under several slightly different definitions in the
literature.  In this paper we establish the actual relationships
between the main four such alternative definitions: namely, the
original galled trees, level-1 networks, nested networks with nesting
depth 1, and evolutionary networks with arc-disjoint reticulation
cycles.
\end{abstract}

\section{Introduction}

The extension of traditional phylogenetic methods and tools to deal
with reticulate evolution is hindered by the computational complexity
of phylogenetic reconstruction.  Several techniques such as parsimony
and likelihood have been carried over from phylogenetic trees to
networks~\cite{hein:1990,nakhleh.ea:bioinfo06,nakhleh.ea:bioinfo07,strimmer.ea:2000},
but when it comes to exact methods for phylogenetic reconstruction,
the hardness of reconstructing an evolutionary network with as few
reticulations as possible for a given set of sequences was soon
established~\cite{gusfield.ea:csb:2003,wang.ea:2000,wang.ea:2001}.

Under suitable constraints on reticulation cycles, however, the latter
problem can be solved in polynomial time.  For instance, the so-called
\emph{galled trees} \cite{gusfield.ea:csb:2003}, evolutionary networks
with disjoint reticulation cycles, can be reconstructed in time
polynomial in the size of the sequences and, when they exist, contain
the smallest possible number of reticulations that explain the
evolutionary history of the given set of sequences under mutation and
recombination, with the assumption of no back or recurrent
mutations~\cite{gusfield:jcss:2005,gusfield.ea:csb:2003,gusfield.ea:fine.structure:2004,gusfield.ea:galled.trees:2004}.

The ``disjoint reticulation cycles" condition for galled trees has appeared
several times and in different guises in the literature.  Up to our
knowledge, it was first introduced as the \emph{condition C2} in
\cite{MWL1998,wang.ea:2000} and in the definition of \emph{perfect
phylogenetic networks with recombination} in \cite{wang.ea:2001}.
Gusfield \textsl{et al}'s original definition of (the topology of a)
\emph{galled tree} is as a rooted DAG with all hybrid nodes of
in-degree 2 and without nodes belonging to two reticulation cycles.
In this definition, the restriction to hybrid nodes of in-degree 2 is
imposed by their semantics: they represent very specific recombination
operations of pairs of sequences.  Although the original definition of
galled tree imposes that reticulation cycles are disjoint at the level
of nodes, it has been realized that their combinatorial analysis also
works if they are only required to be disjoint at the level of edges
\cite{HusonKloepper2007}: lacking of a specific term for the resulting
networks, we shall call them here \emph{weakly galled trees}, to
distinguish them from the original galled trees.

Soon later, Jansson and Sung \cite{JanssonSung2004a,JanssonSung2008}
introduced the \emph{nested networks} and they claimed that the nested
networks of \emph{nesting depth 1} (which, for simplicity, we shall
abbreviate henceforth as \emph{1-nested networks}) were the same as the
galled trees in the sense of \cite{gusfield.ea:csb:2003}.  1-nested
networks, as defined in \textsl{loc.  cit.}, are bijectively
leaf-labelled rooted DAGs with some restrictions on the degrees of the
nodes (namely, the tree nodes, including the root, have out-degree 0
or 2, and the hybrid nodes have in-degree 2 and out-degree 0, 1 or 2)
and where no node is intermediate in reticulation cycles for different
hybrid nodes.  Also, Jansson, Sung and collaborators defined
\emph{level-k} networks \cite{CJSS2005,jansson.sung:2006} as
bijectively leaf-labelled rooted DAGs, with the same restrictions on
the degrees of the nodes as in 1-nested networks, and where each
biconnected subgraph contains at most $k$ hybrid nodes, and they also
claimed that level-1 networks were the galled trees.  The restrictions
on the degrees of the nodes in 1-nested or level-1 networks have no
semantical meaning, being necessary only to guarantee that the
reconstruction algorithms proposed in those papers run in
polynomial-time.  Thus, it is plausible that, in the future, these
restrictions can be relaxed, if new algorithms using other kinds of
data produce in polynomial time networks satisfying the defining
conditions of 1-nested or level-1 networks, but with tree nodes of
out-degree greater than 2 or hybrid nodes of in-degree greater than 2
(corresponding to combinations of mutations or recombinations,
respectively, where the order of the events cannot be ascertained
exactly \cite{semple:07}).  For this reason, we do not include in our
definitions of 1-nested and level-1 networks any restriction on the
nodes' degrees.

Under one name or the other, galled trees have fostered much research
on phylogenetic network
structure~\cite{gusfield.ea:jcb:2007,semple.steel:2006}, tight bounds
on the number of
reticulations~\cite{gusfield.hickerson.eddhu:2007,song.wu.gusfield:2005},
and reconstruction
algorithms~\cite{CJLY2006,gusfield.ea:csb:2003,jansson.ea:sicomp:2006,jansson.sung:2006,NWLS2005,van.iersel.ea:2008}.
The goal of this paper is to study the actual relationship among
galled trees, weakly galled trees, 1-nested networks, and level-1
networks, establishing in particular to which extent 1-nested networks
and level-1 networks are actually galled trees.  Among other things,
we prove that, under the ``hybrid nodes of in-degree 2" restriction,
the 1-nested networks are exactly the weakly galled trees, and that
the class they define strictly contains the level-1 networks, which,
on their turn, are strictly more general than the galled trees.
However, in the fully resolved case, all four definitions describe
exactly the same networks.

\section{Preliminaries}

By an \emph{evolutionary network} on a set $S$ of \emph{taxa} we simply
mean a rooted DAG with its leaves bijectively labeled in $S$.

A \emph{tree node} of an evolutionary network $N=(V,E)$ is a node of
in-degree at most 1, and a \emph{hybrid node} is a node of in-degree
at least 2.  A \emph{tree arc} (respectively, a \emph{hybridization
arc}) is an arc with head a tree node (respectively, a hybrid node).
A node $v\in V$ is a \emph{child} of $u\in V$ if $(u,v)\in E$; we also
say in this case that $u$ is a \emph{parent} of $v$.  We denote by
$u\pathgr v$ any path in $N$ with origin $u$ and end $v$.  Whenever
there exists a path $u\pathgr v$, we shall say that $v$ is a
\emph{descendant} of $u$ and also that $u$ is an \emph{ancestor} of
$v$.  A path $u\pathgr v$ is \emph{non-trivial} when $u\neq v$: in
this case, we say that $v$ is a \emph{proper descendant} of $u$ and
that $u$ is a \emph{proper ancestor} of $v$.  A \emph{minimal common
ancestor} (\emph{mca}, for short) of a pair of nodes $u,v$ is a common
ancestor of $u$ and $v$ that is not a proper ancestor of any other
common ancestor of them.

We shall say that an evolutionary network is \emph{2-hybrid} when its
hybrid nodes have in-degree 2, \emph{hybrid-1} when its hybrid nodes
have out-degree 1, \emph{semibinary} when its hybrid nodes have
in-degree 2 and out-degree 1, and \emph{binary}, or \emph{fully
resolved}, when it is semibinary and its internal tree nodes have
out-degree 2.

Two paths in an evolutionary network are said to be \emph{internally
disjoint} when they have disjoint sets of intermediate nodes.  A
\emph{reticulation cycle} for a hybrid node $h$ is a pair of
internally disjoint paths ending in $h$ and with the same origin.
Each one of the paths forming a reticulation cycle for a node $h$ is
called generically a \emph{merge path} for $h$, their common origin is
called the \emph{split node} of the reticulation cycle, and the hybrid
node $h$, the \emph{end} of the reticulation cycle.  The
\emph{intermediate nodes} of a reticulation cycle are the intermediate
nodes of the merge paths forming it.

\begin{remark}
Let $h$ be a hybrid node and let $u$ and $v$ be two different proper
ancestors of it such that the paths $u\pathgr h$ and $v\pathgr h$ have
only their end $h$ in common.  Let $w$ be a mca of $u$ and $v$.  If
$w\neq u,v$ (which in particular implies that $u$ and $v$ are not
connected by a path), then the paths $w\pathgr u$ and $w\pathgr v$
have only their origin in common, and then the concatenations
$w\pathgr u \pathgr h$ and $w\pathgr v\pathgr h$ define a reticulation
cycle.
If, on the contrary, $w$ is one of the nodes $u,v$, say $w=u$, then
$u$ is an ancestor of $v$ and the only mca of $u$ and $v$.  In this
case there are two possibilities.  If there exists some path $u\pathgr
v$ internally disjoint from $u\pathgr h$, then the paths $u\pathgr h$
and $u\pathgr v\pathgr h$ define a reticulation cycle for $h$, with
split node $u$.
But if there does not exist any path $u\pathgr v$ internally disjoint
from $u\pathgr h$, and if $w$ is the {last} node in the path $u\pathgr
h$ that is an ancestor of $v$, then the subpath $w\pathgr h$ of
$u\pathgr h$ and the path $w\pathgr v\pathgr h$ form a reticulation
cycle, with split node $w$.
\end{remark}

A straightforward consequence of this observation is the following
lemma, which will be used several times in the next sections.

\begin{lemma}\label{fact2}
If an evolutionary network contains a hybrid node $h$ and two
non-trivial paths $v_1\pathgr h$ and $v_2\pathgr h$ with only their
end $h$ in common, then either $v_1$ and $v_2$ are intermediate nodes
in a reticulation cycle for $h$, or one of the nodes $v_1,v_2$ is
intermediate in a reticulation cycle for $h$ whose split node is a
descendant of the other node.  \qed
\end{lemma}
%

By restricting the possible type of intersections between reticulation
cycles, we obtain different types of evolutionary networks:
\begin{itemize}
\item An evolutionary network is a \emph{galled tree}
\cite{gusfield.ea:csb:2003} when every pair of reticulation cycles
have disjoint sets of nodes.

\item An evolutionary network is a \emph{weakly galled tree} when
every pair of reticulation cycles have disjoint sets of arcs.

\item An evolutionary network is \emph{1-nested} when every pair of
reticulation cycles with different ends have disjoint sets of
intermediate nodes.
\end{itemize}

The last definition deserves some context.  Jansson and Sung
\cite{JanssonSung2004a,JanssonSung2008} define an evolutionary network
to be \emph{nested} when, for every pair of hybrid nodes $h_1$, $h_2$,
one of the following three conditions holds:
\begin{itemize}
\item Every merge path for $h_1$ and every merge path for $h_2$ are
internally disjoint.  \item Every merge path for $h_1$ is a subpath of
some merge path for $h_2$.  \item Every merge path for $h_2$ is a
subpath of some merge path for $h_1$.
\end{itemize}
Then, they define a nested evolutionary network to have \emph{nesting
depth $k$} when every node is an intermediate node of reticulation
cycles for at most $k$ hybrid nodes.  Now, notice that the nesting
depth 1 condition implies the nested condition (because every pair of
merge paths for different hybrid nodes will be internally disjoint),
and therefore the nested networks with nesting depth 1 are exactly the
evolutionary networks where no node is intermediate in reticulation
cycles for more than one hybrid node, which are the networks we have
dubbed 1-nested.

A subgraph of an undirected graph is \emph{biconnected} when it is
connected and it remains connected if we remove any node and all edges
incident to it.  A subgraph of an evolutionary network $N$ is said to
be \emph{biconnected} when it is so in the undirected graph associated
to $N$.  Every arc in an evolutionary network is a biconnected
subgraph.  Every reticulation cycle also induces a biconnected
subgraph.

\begin{remark}\label{muc}
If a pair of nodes in an evolutionary network $N$ belong to a
biconnected subgraph with more than 2 nodes, then they must belong to
some {minimal}\footnote{By a \emph{minimal cycle}
$(v_0,v_{1},\ldots,v_{k},v_0)$ we mean a cycle such that the nodes
$v_0,v_1,\ldots,v_{k}$ are pairwise different.} cycle contained in the
corresponding biconnected subgraph of the undirected graph associated
to $N$.  This minimal cycle will correspond in $N$ to a sequence of
$2k$ (directed) different paths
$$
v_1\pathgr h_1, v_1\pathgr h_2, v_2\pathgr h_2, v_2\pathgr h_3,\ldots,
v_k\pathgr h_k, v_k\pathgr h_1
$$
where $h_1,\ldots,h_k$ are pairwise different hybrid nodes,
$v_1,\ldots,v_k$ are pairwise different nodes, and the only possible
intersection between a pair of such paths is to share the origin or
the end in the way indicated by the notations.
To simplify the language, we shall call such a sequence of paths in
$N$ a \emph{minimal undirected cycle}.
\end{remark}

An evolutionary network is \emph{level-$k$}
\cite{CJSS2005,jansson.sung:2006,IKM2009} when no biconnected subgraph
of it contains more than $k$ hybrid nodes.  Thus, a \emph{level-1
network} is an evolutionary network where no biconnected subgraph
contains more than 1 hybrid node.  In particular, the minimal
undirected cycles in a level-1 network are reticulation cycles with
split node of tree type and no hybrid intermediate node.

\begin{figure}[htb]
\begin{center}
\begin{tikzpicture}[thick,>=stealth,scale=0.4]
\draw (0,0) node[tre] (1) {}; \etq 1
\draw (2,0) node[tre] (2) {}; \etq 2
\draw (4,0) node[tre] (3) {}; \etq 3
\draw (6,0) node[tre] (4) {}; \etq 4
\draw (8,0) node[tre] (5) {}; \etq 5
\draw (6,2) node[hyb] (B) {};  
\draw (2,4) node[hyb] (A) {};  
\draw (4,4) node[tre] (c) {};  
\draw (8,4) node[tre] (d) {};  
\draw (0,6) node[tre] (a) {};  
\draw (6,6) node[tre] (b) {};  
\draw (4,8) node[tre] (r) {};  
\draw[->](r)--(a);
\draw [->](r)--(b);
\draw [->](a)--(A);
\draw [->](b)--(A);
\draw [->](b)--(c);
\draw [->](b)--(d);
\draw [->](a)--(1);
\draw [->](A)--(2);
\draw [->](c)--(3);
\draw [->](c)--(B);
\draw [->](B)--(4);
\draw [->](d)--(B);
\draw [->](d)--(5);
\draw (4,-2) node {(a)};
\end{tikzpicture}
\qquad
\begin{tikzpicture}[thick,>=stealth,scale=0.4]
\draw (0,0) node[tre] (1) {}; \etq 1
\draw (2,0) node[tre] (2) {}; \etq 2
\draw (4,0) node[tre] (3) {}; \etq 3
\draw (6,0) node[tre] (4) {}; \etq 4
\draw (8,0) node[tre] (5) {}; \etq 5
\draw (4,2) node[hyb] (B) {};  
\draw (2,4) node[tre] (c) {};  
\draw (6,4) node[tre] (d) {};  
\draw (4,6) node[hyb] (A) {};   
\draw (1,8) node[tre] (a) {};  
\draw (7,8) node[tre] (b) {};  
\draw (4,9) node[tre] (r) {};  
\draw[->](r)--(a);
\draw [->](r)--(b);
\draw [->](a)--(A);
\draw [->](b)--(A);
\draw [->](a)--(1);
\draw [->](b)--(5);
\draw [->](A)--(c);
\draw [->](A)--(d);
\draw [->](c)--(B);
\draw [->](c)--(2);
\draw [->](d)--(B);
\draw [->](d)--(4);
\draw [->](B)--(3);
\draw (4,-2) node {(b)};
\end{tikzpicture}
\end{center}
\caption{\label{fig:3} Two  weakly galled trees that are not galled trees.}
\end{figure}
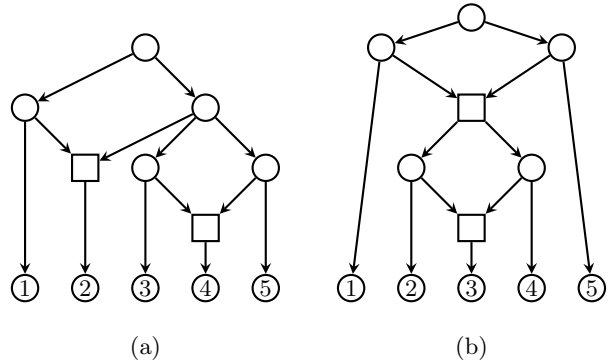

\begin{remark}\label{rem:obvi}
It is clear from the definitions that every galled tree is a weakly
galled tree.  The converse implication is false: see Fig.~\ref{fig:3}.
\end{remark}

Galled trees were originally defined as being 2-hybrid, because their
hybrid nodes represented recombinations of pairs of sequences
\cite{gusfield.ea:csb:2003}.  Nevertheless, it turns out that the
condition of having arc-disjoint reticulation cycles implies that all
hybrid nodes must have in-degree 2.

\begin{lemma}\label{lem:gt->2h}
Every weakly galled tree (and hence every galled tree) is 2-hybrid. 
\end{lemma}

\begin{proof}
If an evolutionary network $N$ contains some hybrid node $h$ with three
different parents $a,b,c$, then it contains some reticulation cycle
for $h$ with merge paths ending in $(a,h)$ and $(b,h)$, and some other
reticulation cycle with merge paths ending in $(b,h)$ and $(c,h)$.
These reticulation cycles share the arc $(b,h)$, which shows that $N$
is not a weakly galled tree.
\end{proof}

\begin{figure}[htb]
\begin{center}
\begin{tikzpicture}[thick,>=stealth,scale=0.5]
\draw (0,0) node[tre] (1) {}; \etq 1
\draw (2,0) node[tre] (2) {}; \etq 2
\draw (4,0) node[tre] (3) {}; \etq 3
\draw (6,0) node[tre] (4) {}; \etq 4
\draw (4,2) node[hyb] (h) {};  
\draw (1,4) node[tre] (a) {};  
\draw (3,4) node[tre] (b) {};  
\draw (5,4) node[tre] (c) {};  
\draw (3,6) node[tre] (r) {};  
\draw[->](r)--(a);
\draw [->](r)--(b);
\draw [->](r)--(c);
\draw [->](a)--(1);
\draw [->](a)--(h);
\draw [->](b)--(2);
\draw [->](b)--(h);
\draw [->](c)--(4);
\draw [->](c)--(h);
\draw [->](h)--(3);
\end{tikzpicture}
\end{center}
\caption{\label{fig:0.2} An evolutionary network that is 1-nested and
level-1 but not 2-hybrid.}
\end{figure}
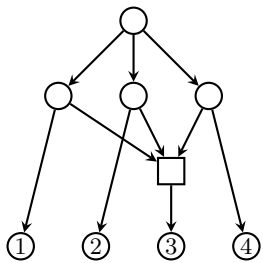

Unlike galled trees, 1-nested and level-1 evolutionary networks need
not be 2-hybrid: see Fig.~\ref{fig:0.2}.

\section{Results for Arbitrary Networks}

In this section we investigate the relationship between level-1
networks and 1-nested networks when no restriction on the in-degrees
of hybrid nodes in the networks is imposed.

\begin{proposition}\label{l1->1n}
Every level-1 network is 1-nested.
\end{proposition}

\begin{proof}
Let $N$ be a level-1 evolutionary network and assume that it contains
some node $v$ that is intermediate in reticulation cycles for two
different hybrid nodes $h_1$ and $h_2$.  The node $v$ must be of tree
type, because otherwise the reticulation cycles of $h_1$ and $h_2$
would be biconnected subgraphs of $N$ with more than one hybrid node,
which is forbidden in level-1 networks.  Let $w$ be the only parent of
$v$ ($v$ cannot be the root, because it is intermediate in
reticulation cycles).  Then the arc $(w,v)$ must belong to the
reticulation cycles for $h_1$ and $h_2$ that contain $v$.  This
implies that the union of these two reticulation cycles is a
biconnected subgraph of $N$, against the assumption that $N$ is
level-1.
\end{proof}

The converse implication is in general false: network (b) in
Fig.~\ref{fig:3} is 1-nested, but not level-1.  Actually, that
counterexample captures the only pathology that can prevent a 1-nested
network from being level-1, as Theorem \ref{1n->l1} below shows.  To
prove it, we shall use the following lemma.

\begin{lemma}\label{lem:2h}
In a 1-nested network, no reticulation cycle contains an intermediate
hybrid node.
\end{lemma}

\begin{proof}
Let $N$ be a 1-nested network, and assume that a hybrid node $h_1$ is
intermediate in a reticulation cycle $C$ for a hybrid node $h_2$, and
let $P:h_1\pathgr h_2$ be the subpath of the corresponding merge path.
Let $u$ be the split node of the reticulation cycle $C$, and let $P_1:
u\pathgr h_1\pathgr h_2$ and $P_2: u\pathgr h_2$ be the merge paths of
this reticulation cycle.  Let now $v_2$ be a parent of $h_1$ that is
not the node preceding $h_1$ in the path $P_1$.

Assume that $v_2$ belongs to the path $P_1$.  In this case the subpath
$v_2\pathgr h_1$ of $P_1$ and the arc $(v_2,h_1)$ form a reticulation
cycle $C'$ for $h_1$, and the node preceding $h_1$ in the path $P_1$
will be intermediate in the reticulation cycles $C'$ for $h_1$ and $C$
for $h_2$, against the assumption that $N$ is 1-nested.

Assume now that $v_2$ belongs to the path $P_2$: since we have already
discarded the possibility that $u=v_2$, 
and $h_2 \neq v_2$ because $N$ is acyclic,
it will be intermediate in
$P_2$.  In this case, the subpath $u\pathgr h_1$ of $P_1$ and the
concatenation $u\pathgr v_2\to h_1$ of the subpath $u\pathgr v_2$ of
$P_2$ and the arc $(v_2,h_1)$ form a reticulation cycle $C'$ for
$h_1$, and $v_2$ is intermediate in the reticulation cycles $C'$ for
$h_1$ and $C$ for $h_2$, which again contradicts the assumption that
$N$ is 1-nested.

So, $v_2$ does not belong to the paths $P_1$ or $P_2$.  Let $v$ be a
mca of $v_2$ and $u$.  We must distinguish now several cases, in all
of which we obtain a node that is intermediate in reticulation cycles
for $h_1$ and $h_2$, contradicting the assumption that $N$ is
1-nested:

\begin{itemize}
\item If $v\neq v_2,u$, then it defines a reticulation cycle $C'$ for
$h_1$ with split node $v$ and merge paths $P_3: v\pathgr v_2\to h_1$
and $P_4: v\pathgr u\pathgr h_1$.  In this case, since $v_2$ is
neither an ancestor nor a descendant of $u$ (because $v\neq v_2,u$),
the paths obtained by concatenating, on the one hand, the paths $P_3$
and $P$ and, on the other hand, the subpath $v\pathgr u$ of $P_4$ and
the path $P_2$, form a new reticulation cycle $C''$ for $h_2$, with
split node $v$.  Therefore, $v_2$ is an intermediate node in the
reticulation cycles $C'$ for $h_1$ and $C''$ for $h_2$.

\item If $v=v_2$, then the arc $(v_2,h_1)$ and the path
$P_3:v_2\pathgr u\pathgr h_1$ form a reticulation cycle $C'$ for
$h_1$.  In this case, $v_2$ is also the split node of a reticulation
cycle $C''$ for $h_2$, with merge paths on the one hand the
concatenation of the arc $(v_2,h_1)$ and the path $P$ and, on the
other hand, the concatenation of the subpath $v_2\pathgr u$ of $P_3$
and the path $P_2$.  In this way, $u$ turns out to be intermediate in
the reticulation cycles $C'$ for $h_1$ and $C''$ for $h_2$.

\item If $v=u$ and $v_2$ is not a descendant of any intermediate node
in the subpath $u\pathgr h_1$ of $P_1$, then $u$ is the split node of
a reticulation cycle $C'$ for $h_1$, with merge paths $P_3: u\pathgr
h_1$ (the corresponding subpath of $P_1$) and $P_4: u\pathgr v_2\to
h_1$.  Now, the subpath $u\pathgr v_2$ of $P_4$ may have more nodes in
common with $P_2:u\pathgr h_2$ than the origin.  Let $w$ be the last
node in $P_2$ that appears in the path $u\pathgr v_2$.  Since $w\neq
v_2$ (because we already know that $v_2$ does not belong to $P_2$),
the subpath $w\pathgr h_2$ of $P_2$ and the concatenation of the
subpath $w\pathgr v_2\to h_1$ of $P_4$ with $P$ define a reticulation
cycle $C''$ for $h_2$, and $v_2$ is intermediate in this reticulation
cycle for $h_2$ as well as in the reticulation cycle $C'$ for $h_1$.
 
\item If $v=u$ but $v_2$ is a proper descendant of some intermediate
node in the subpath $u\pathgr h_1$ of $P_1$, then let $w$ be the last
intermediate node in $u\pathgr h_1$ with this property: in this case,
$w$ is the split node of a reticulation cycle $C'$ consisting of the
merge paths $w\pathgr v_2\to h_1$ and the subpath $w\pathgr h_1$ of
$P_1$.  Now, the paths $w\pathgr v_2$ and $P_2$ may have some node in
common, which leads to two possibilities:
\begin{itemize}
\item If the paths $w\pathgr v_2$ and $P_2$ are disjoint, then the
path $P_2$ and the concatenation of the subpath $u\pathgr w$ of $P_1$
with the path $w\pathgr v_2\to h_1$ followed by $P$ form a
reticulation cycle $C''$ for $h_2$ that has $v_2$ as an intermediate
node, and $v_2$ was already an intermediate node of the reticulation
cycle $C'$ for $h_1$.

\item If the paths $w\pathgr v_2$ and $P_2$ are not disjoint, let $w'$
be the last node in $P_2$ that also belongs to $w\pathgr v_2$.  Since
$w'\neq v_2$, because $v_2$ does not belong to $P_2$, the subpath
$w'\pathgr h_2$ 
of $P_2$
and the concatenation of the subpath $w'\pathgr v_2$
of $w\pathgr v_2$ with the arc $(v_2,h_1)$ and the path $P$ yields a
reticulation cycle $C''$ for $h_2$ with split node $w'$.  Then, $v_2$
is intermediate in this reticulation cycle for $h_2$ as well as in the
reticulation cycle $C'$ for $h_1$.
\end{itemize}
\end{itemize}
Thus, all possible situations arising when a hybrid node is intermediate in a reticulation cycle lead to a contradiction in 1-nested networks.
\end{proof}

\begin{theorem}\label{1n->l1}
The level-1 networks are exactly the 1-nested networks without hybrid
split nodes.
\end{theorem}

\begin{proof}
Every level-1 network is 1-nested by Proposition \ref{l1->1n}, and it
has no hybrid split node, because a reticulation cycle with hybrid
split node induces a biconnected subgraph with more than one hybrid node.

As far as the converse implication goes, let $N$ be a 1-nested network
where no hybrid node is the split node of any reticulation cycle.  Let
us assume that $N$ contains some biconnected subgraph, and in
particular some minimal undirected cycle in the sense of Remark
\ref{muc}, with more than one hybrid node, and let us see that this
leads to a contradiction.  This will prove that $N$ is level-1.

The minimal undirected cycle of $N$ with at least two hybrid nodes
cannot be a reticulation cycle, because no reticulation cycle in $N$
contains any hybrid node other than its end: the split node of a
reticulation cycle in $N$ cannot be hybrid by assumption, and no
intermediate node of a reticulation cycle in $N$ can be hybrid by
Lemma \ref{lem:2h}.  Therefore, this minimal undirected cycle will
consist of $2k$ paths, with $k\geq 2$,
$$
v_1\pathgr h_1, v_1\pathgr h_2, v_2\pathgr h_2, v_2\pathgr h_3,v_3\pathgr h_3, v_3\pathgr h_4,\ldots,
v_k\pathgr h_k, v_k\pathgr h_1.
$$
Applying Lemma \ref{fact2} to the paths $v_1\pathgr h_1$ and
$v_k\pathgr h_1$, we obtain that at least one of the nodes $v_1$ or
$v_k$ is an intermediate node in a reticulation cycle for $h_1$.
Assume that $v_1$ has this property (if $v_1$ was not intermediate in
a reticulation cycle for $h_1$, then $v_k$ would be so, and we would
traverse the cycle in the reverse sense).  Then, applying Lemma
\ref{fact2} to the paths $v_1\pathgr h_2$ and $v_2\pathgr h_2$, and
recalling that $v_1$ cannot be an intermediate node of a reticulation
cycle for $h_2$ (because it is already so for $h_1$), we deduce that
$v_2$ is an intermediate node of a reticulation cycle for $h_2$ and a
descendant of $v_1$.  Now, applying Lemma \ref{fact2} to the paths
$v_2\pathgr h_3$ and $v_3\pathgr h_3$, and since $v_2$ cannot be an
intermediate node of a reticulation cycle for $h_3$, we deduce that
$v_3$ is an intermediate node of a reticulation cycle for $h_3$ and a
descendant of $v_2$, and hence of $v_1$.  Repeating this process, when
we reach $v_k$ we obtain that it must be an intermediate node of a
reticulation cycle for $h_k$ and a descendant of $v_1$.  But then,
$v_k$ cannot be intermediate in the first reticulation cycle for
$h_1$, and therefore $v_1$ must be a descendant of $v_k$, which yields
a contradiction.
\end{proof}

One possible way to forbid hybrid split nodes is to impose that the
hybrid nodes have out-degree 1.  This is usually done when hybrid
nodes represent reticulation events (like hybridizations,
recombinations, or horizontal gene transfers): the only child of a
hybrid node represents then the species resulting from the
reticulation event.

\begin{corollary}\label{h1+1n->l1}
Every 1-nested hybrid-1 network is level-1.
\end{corollary}

In \cite[Lem.~3]{cardona.ea:07b} we proved that galled trees without out-degree
1 tree nodes are
\emph{tree-child}, that is, that every internal node in a galled tree
has some child of tree type.  A suitable modification of the argument
used therein proves the following result.

\begin{proposition}
Every 1-nested (and, hence, every level-1) network without out-degree
1 tree nodes is tree-child.
\end{proposition}

\begin{proof}
Let $N$ be a 1-nested network and let $v$ be an internal node.  There
are two cases to consider.

On the one hand, if $v$ has only one child, then this child is a tree
node.  Indeed, by assumption, if $v$ has only one child, then $v$ must
be hybrid.  But then, since 1-nested networks cannot contain hybrid
nodes that are intermediate in reticulation cycles (Lemma
\ref{lem:2h}), if the child $w$ of $v$ is hybrid, $v$ must be the
split node of a reticulation cycle for $w$, and hence it must have at
least two children.

On the other hand, if $v$ has more than one child, then some child is
a tree node.  Indeed, assume that $v$ has two hybrid children $h_1$
and $h_2$.  Then, the 1-nested condition entails that $v$ cannot be
intermediate in reticulation cycles for both of them, and therefore it
must be the split node of a reticulation cycle for at least one of
them, say for $h_1$.  But then some other child of $v$ must be
intermediate in this reticulation cycle, and this child must be of
tree type, again by Lemma \ref{lem:2h}.
 \end{proof}

\section{Results for 2-Hybrid Networks}

Let us consider now the case when hybrid nodes have in-degree 2, in
which case we can include galled and weakly galled trees in our
discussion.

\begin{lemma}\label{lem:gt}
In a 2-hybrid 1-nested network, each hybrid node is the end of only
one reticulation cycle.
\end{lemma}

\begin{proof}
Let $N$ be a 2-hybrid 1-nested network, and assume that it contains
two reticulation cycles $C,C'$ for a hybrid node $h$, with split
nodes $w_1$ and $w_2$, respectively.  Let $(v_1,h),(v_2,h)$ be the
pair of arcs with head $h$.  Then, in each reticulation cycle for $h$,
one merge path ends in $(v_1,h)$ and the other in $(v_2,h)$.  Let
$P_{1,1}: w_1\pathgr v_1\to h$ and $P_{2,1}:w_2\pathgr v_1\to h$ be
the merge paths of $C$ and $C'$, respectively, ending in $(v_1,h)$,
and let $u$ be the first node in $P_{1,1}$ and $P_{2,1}$ such that the
subpaths $u\pathgr h$ of $P_{1,1}$ and $P_{2,1}$ are the same.  If $u$
is intermediate in both merge paths, this means that is has two
different parents (one in each path) and therefore that it is hybrid,
which contradicts 
Lemma
\ref{lem:2h}.  Therefore there are three
possibilities: either $u=w_1=w_2$, and then the paths $P_{1,1}$ and
$P_{2,1}$ are the same, or $u=w_1$ and it is intermediate in
$P_{2,1}$, and then $P_{1,1}$ is a subpath of $P_{2,1}$, or $u=w_2$
and it is intermediate in $P_{1,1}$, and then $P_{2,1}$ is a subpath
of $P_{1,1}$.  In particular, $w_1$ and $w_2$ are either equal or
connected by a piece of a merge path.

Using the same reasoning, we conclude that, if $P_{1,2}: w_1\pathgr
v_2\to h$ and $P_{2,2}:w_2\pathgr v_2\to h$ are the merge paths of
$C$ and $C'$, respectively, ending in $(v_2,h)$, then either
$P_{1,2}=P_{2,2}$, or $P_{1,2}$ is a subpath of $P_{2,2}$, or
$P_{2,2}$ is a subpath of $P_{1,2}$.  Now all combinations yield to
contradictions: if $w_1=w_2$, then $P_{1,1}=P_{2,1}$ and
$P_{1,2}=P_{2,2}$ and hence $C=C'$; if $w_1$ is a proper descendant
of $w_2$, then $w_1$ is intermediate in $P_{2,1}$ and $P_{2,2}$, and
then these paths are not internally disjoint; and if $w_2$ is a proper
descendant of $w_1$, then $w_2$ is intermediate in $P_{1,1}$ and
$P_{1,2}$, and these paths are not internally disjoint.
\end{proof}

\begin{proposition}\label{2h1n<->wgt}
A 2-hybrid network is 1-nested if, and only if, it is a weakly galled tree.
\end{proposition}

\begin{proof}
Let $N$ be a 2-hybrid 1-nested network, and assume that two
reticulation cycles $C,C'$ share one arc $(u,v)$; by the previous
lemma, these reticulation cycles have different ends, say $h$ and
$h'$, respectively.  Now, neither $u$ nor $v$ are intermediate in both
cycles, because it would contradict the 1-nested condition.  Therefore
$u$ must be the split node of one of the cycles, say $C$, and $v$ must
be $h$ or $h'$: but if 
$v=h$, then it is intermediate in $C'$, and if
$v=h'$, then it is intermediate in $C$, and neither one thing nor the
other is possible, by Lemma \ref{lem:2h}.  This shows that $N$ is a
weakly galled tree.

As far as the converse implication goes, let $N$ be a weakly galled
tree and assume that two reticulation cycles $C$ and $C'$ of $N$ share
an intermediate node $v$.  If $v$ 
were 
a hybrid node, then $C$ would
share an arc with some reticulation cycle with end $v$
(both arcs ending in $v$ belong to any reticulation cycle for $v$, and one of them would belong to $C$), which would
contradict the weakly galled tree condition.
Then, $v$ must be a tree
node.  But in this case the only arc with head $v$ must belong to $C$
and $C'$, and hence these reticulation cycles share an arc, which is
again impossible.
\end{proof}

\begin{corollary}\label{gt->l1}
Every galled tree is a level-1 network.
\end{corollary}

\begin{proof}
Every galled tree is a weakly galled tree, and hence 1-nested by
Proposition \ref{2h1n<->wgt}, and it cannot have any hybrid split
node, because different reticulation cycles cannot have any node in
common.  Then, Theorem \ref{1n->l1} applies.
\end{proof}

\begin{corollary}\label{cor:semibin}
In the semibinary case, level-1 networks, 1-nested networks and weakly
galled trees are the same.
\end{corollary}

\begin{proof}
It is a direct consequence of Propositions \ref{l1->1n} and
\ref{2h1n<->wgt}, and Corollary \ref{h1+1n->l1}.
\end{proof}

\begin{remark}
Not every 2-hybrid 1-nested network is level-1: see network (b) in
Fig.~\ref{fig:3}.  And not every semibinary level-1 network is a
galled tree: see network (a) in Fig.~\ref{fig:3}.
\end{remark}

\begin{proposition}
In the binary case, level-1 networks, 1-nested networks, weakly galled
trees, and galled trees are the same.
\end{proposition}

\begin{proof}
By Corollaries \ref{gt->l1} and \ref{cor:semibin}, it is enough to
prove that every binary 1-nested network is a galled tree.  So, let
$N$ be a binary 1-nested network, and assume that two reticulation
cycles $C,C'$ share one node.  By Lemma \ref{lem:gt} we know that $C$
and $C'$ have different hybrid ends, say $h$ and $h'$.  In particular,
they do not share their hybrid end.  Moreover, the node they share
cannot be intermediate in both cycles either, because $N$ is 1-nested.
Let us see that all the other possibilities also lead to a
contradiction:
\begin{itemize}
\item The hybrid end of one of the cycles cannot be the split node of
the other, because split nodes cannot have out-degree 1 and hybrid
nodes in binary networks have out-degree 1.

\item The hybrid end of one of the cycles cannot be intermediate in
the other, because of Lemma \ref{lem:2h}.

\item If the split node of one of the cycles, say $C$, belongs to the
other cycle, then (since it cannot be its hybrid end), one of its
children in $C$ must be its child in $C'$, otherwise the split node
would have out-degree 3.  Now, this shared child of the split node of
$C$ cannot be the hybrid end of $C$ or $C'$ (if it 
were 
the hybrid end
of one of the cycles, it would be an intermediate hybrid node of the
other cycle, against Lemma \ref{lem:2h}).  Therefore, the shared child
of the split node of $C$ will be intermediate in $C$ and in $C'$,
which is prevented by the 1-nested condition.
\end{itemize}
\end{proof}

\section{Conclusion}
In this paper we have established the actual relationships between the
classes of galled trees, weakly galled trees, level-1 networks, and
1-nested networks.  Our main results are summarized as follows:
\begin{enumerate}[(a)]
\item For arbitrary networks,
$$
\mbox{level-1}\Longrightarrow \mbox{1-nested}
$$

\item For hybrid-1 networks,
$$
\mbox{level-1}\Longleftrightarrow \mbox{1-nested}
$$

\item For 2-hybrid networks,
$$
\mbox{galled tree}\Longrightarrow \mbox{level-1}\Longrightarrow
\mbox{1-nested} \Longleftrightarrow \mbox{weakly galled tree}
$$

\item For semibinary networks,
$$
\mbox{galled tree}\Longrightarrow \mbox{level-1}\Longleftrightarrow
\mbox{1-nested} \Longleftrightarrow \mbox{weakly galled tree}
$$

\item For binary networks,
$$
\mbox{galled tree} \Longleftrightarrow
\mbox{level-1}\Longleftrightarrow \mbox{1-nested} \Longleftrightarrow
\mbox{weakly galled tree}
$$
\end{enumerate}
So, if we restrict ourselves to 2-hybrid networks, we see that the
node-disjoint reticulation cycles condition is the most restrictive
one and that 1-nested networks are the most general, being equal to
those networks with arc-disjoint reticulation cycles.  So, since these networks have the same combinatorial properties as galled trees \cite{HusonKloepper2007}, 
from a formal point of view they are probably the right notion of ``phylogenetic network with isolated reticulation cycles".
However, the
distinction between node-disjoint and arc-disjoint reticulation cycles
is very important in practice, because the assumption of no back or
recurrent mutations entails that all nodes are labeled by different
sequences and then, two arc-disjoint, but not node-disjoint,
reticulation cycles cannot be torn apart by just duplicating any common nodes.

\section*{Acknowledgment}

The research reported in this paper has been partially supported by
the Spanish government and the EU FEDER program under project
MTM2006-07773 COMGRIO. We want to thank Gabriel Cardona, Dan Gusfield,
Jesper Jansson and Merc\`e Llabr\'es for several discussions on the
topic of this paper.

\end{document}